\documentclass[showpacs,preprintnumbers,amsmath,amssymb]{revtex4}

\usepackage{graphicx}

\date{\today}

\begin{document}

\title{Time-dependent perturbation theory 
for vibrational energy relaxation and dephasing in peptides and proteins}

\author{Hiroshi Fujisaki}\email{fujisaki@bu.edu} 
\author{Yong Zhang}\email{zhangy@bu.edu} 
\author{John E. Straub}\email{straub@bu.edu} 
\affiliation{
Department of chemistry,
Boston University, 590 Commonwealth Ave.,
Boston, 02215, Massachusetts, USA
}

\begin{abstract}

Without invoking the Markov approximation, we derive formulas 
for vibrational energy relaxation (VER) and dephasing for an anharmonic 
system oscillator using a time-dependent perturbation theory. 
The system-bath Hamiltonian 
contains more than the third order coupling terms
since we take a normal mode picture as a zeroth order approximation. 
When we invoke the Markov approximation, 
our theory reduces to the Maradudin-Fein formula which is used to 
describe VER properties of glass and proteins. 
When the system anharmonicity and the 
renormalization effect due to the environment vanishes, 
our formulas reduce to 
those derived by Mikami and Okazaki invoking the 
path-integral influence functional method [J.~Chem.~Phys.~{\bf 121}, 10052 (2004)].
We apply our formulas to VER of the amide I mode of a small amino-acide like molecule, 
$N$-methylacetamide, in heavy water.

\end{abstract}


\maketitle

\section{Introduction}




Vibrational energy relaxation (VER) and dephasing are fundamental 
properties of molecular dynamics, energy transfer, and reactivity.
Many experimental and theoretical studies have explored these fundamental
processes in gas phase, the liquid state, and in glasses and biomolecular systems \cite{FS05}.
Though our methodology can be applied to any molecular system, 
we are primarily interested in addressing VER and dephasing in peptides or proteins.
While recent advanced experimental techniques using absorption spectra or
time-resolved spectra can deduce  
the structure and dynamics of such a peptide or protein system, 
theoretical approaches are needed to clarify 
the mechanisms of VER and dephasing underlying the experimental data.

The most standard approach to this problem is through the 
perturbation theory of quantum mechanics 
as initiated by Oxtoby \cite{Oxtoby}.
Recently Hynes's group \cite{Hynes} and Skinner's group \cite{Skinner} 
thoroughly studied the VER and dephasing properties of water 
(their target mode was the OH bond of HOD in heavy water)
using this strategy.
This approach is applicable to 
peptides or proteins as was first illustrated by Straub and coworkers \cite{Straub}.
Derived from this strategy is the use of the 
Maradudin-Fein formula (or its equivalent),
which was pursued by Leitner \cite{Leitner05} 
and Straub and coworkers \cite{FBS05a}.
This formula requires the normal modes of the system and 
the cubic anharmonic coefficients between the normal modes.
This methodology can provide a reasonable account of VER
properties of peptides or proteins, 
but there are several deficiencies:
the most serious one is that it assumes 
the Markov properties of the system, so 
it cannot describe the short time dynamics \cite{FBS05b}.
Another problem is the determination of 
the ``lifetime'' width parameter \cite{FS05,FBS05a,FBS05b}.
We also want to describe the dephasing properties 
of the system, crucial to the interpretation of the 
experimental results; it cannot be directly 
described by the MF formula (but see \cite{Leitner02}).

To meet these goals, we derive the formulas for VER 
and dephasing without assuming the Markov properties, i.e., 
without taking an infinite-time limit.
As a result, we can avoid the annoying ``width parameter'' problem
inherent to the MF approach. 
In this sense, Mikami and Okazaki \cite{MO04} took 
a similar path using the path-integral 
influence functional theory.
We use a simple 
time-dependent perturbation theory of quantum mechanics, 
and derive the VER and dephasing formulas more easily. 
We find there is a difference between our formulas and theirs 
in terms of renormalization of the system Hamiltonian.
Another difference is that our system oscillator is taken to be 
a cubic anharmonic oscillator, whereas their mode is a harmonic
oscillator. This can affect the result when the formulas are applied 
to real systems with strong anharmonicity.

This paper is organized as follows:
In Sec.~\ref{sec:derivation},
we derive the VER and dephasing formulas 
for an anharmonic oscillator (mode)
without assuming the Markov properties. 
In Sec.~\ref{sec:NMA}, we apply our formulas 
to the amide I mode of $N$-methylacetamide in
heavy water, and discuss the numerical results and the 
limitations of our strategy.
In Sec.~\ref{sec:summary}, we summarize the paper.
Several system parameters and coefficients in our formulas 
are defined in the Appendix.

\section{Derivation of the formulas for VER and dephasing}
\label{sec:derivation}

\subsection{System, Bath, and Coupling}

We take our Hamiltonian of a solvated 
peptide or protein to be
\begin{eqnarray}
{\cal H} 
&=& {\cal H}_0 + {\cal V} 
= {\cal H}_S +{\cal H}_B+ {\cal V} 
= {\cal H}_S^{(0)} +{\cal H}_f +{\cal H}_B+ {\cal V},
\\
{\cal V}
&=&
-q_S ({\cal F} -\langle {\cal F} \rangle)+ q_S^2 
({\cal G}-\langle {\cal G} \rangle)
=
-q_S \delta{\cal F}+ q_S^2 \delta{\cal G},
\\
{\cal H}_S^{(0)}
&=& 
\frac{p_S^2}{2} + \frac{\omega_S^2}{2} q_S^2 
-q_S \langle {\cal F} \rangle + q_S^2 \langle {\cal G} \rangle
=
\frac{p_S^2}{2} + \frac{\bar{\omega}_S^2}{2} \bar{q}_S^2 
-\frac{\langle {\cal F} \rangle^2}{2 \bar{\omega}_S^2},
\\
{\cal H}_f
&=& \frac{f}{6} \bar{q}_S^3,
\\
{\cal H}_B
&=&
\sum_{\alpha} 
\left(
\frac{p_{\alpha}^2}{2}+\frac{\omega_{\alpha}^2}{2} q_{\alpha}^2
\right),
\end{eqnarray}
where
\begin{eqnarray}
\bar{\omega}_S
&=& \omega_S \sqrt{1+\frac{2 \langle {\cal G} \rangle}{\omega_S^2}}, 
\label{eq:omegas}
\\
\bar{q}_S
&=&
q_S-\frac{\langle {\cal F} \rangle}{\bar{\omega}_S^2}=q_S-b.
\label{eq:qs}
\end{eqnarray}
${\cal H}_S={\cal H}_S^{(0)}+{\cal H}_f$ 
is the renormalized system Hamiltonian
representing a vibrational mode $q_S$ with cubic anharmonicity $f$,
${\cal H}_B$ the bath Hamiltonian representing 
solvent or environmental degrees of freedom with harmonic 
frequencies $\omega_{\alpha}$,
and ${\cal V}$ the interaction Hamiltonian describing the 
coupling between the system and the bath. 
We have assumed that the interation can be 
Taylor expanded, and we have only included up to the 
second order in $q_S$.
Note that we need to renormalize the system 
to assure that $\langle {\cal V} \rangle =0$ where 
the bracket denotes the bath average throughout this paper.
(For the definition of $\delta {\cal F}$ and $\delta {\cal G}$, 
see Appendix \ref{app:system}.)
This is automatically satisfied in the case of 
bilinear coupling like the Caldeira-Leggett-Zwanzig model \cite{Weiss},
but this is not usually the case. 
The system variable becomes ${\bar q}_S$ instead of $q_S$, 
and the system frequency does ${\bar \omega}_S$ instead of $\omega_S$.
This is similar to previous treatments of the system-bath interaction
in the literature \cite{Skinner,XYK02}.

\subsection{Perturbation theory for VER and dephasing}

Starting from the interaction picture of the von Neumann equation, 
we can expand the density operator for the full system as 
\begin{eqnarray}
\tilde{\rho}(t)
&=& \rho(0)+ \frac{1}{i \hbar} \int_0^t dt' 
[\tilde{\cal V}(t'), \tilde{\rho}(t')]
\nonumber
\\
&=&
\rho(0)+ \frac{1}{i \hbar} \int_0^t dt' 
[\tilde{\cal V}(t'), \rho(0)]
\nonumber
\\
&&
+\frac{1}{(i \hbar)^2} \int_0^t dt' \int_0^{t'} dt''  
[\tilde{\cal V}(t'), [\tilde{\cal V}(t''),\rho(0)]]
+ \cdots
\end{eqnarray}
where 
\begin{equation}
\tilde{\rho}(t) \equiv 
e^{i {\cal H}_0 t/\hbar} \rho(t) e^{-i {\cal H}_0 t/\hbar},
\hspace{1cm}
\tilde{\cal V}(t) \equiv 
e^{i {\cal H}_0 t/\hbar} {\cal V} e^{-i {\cal H}_0 t/\hbar}.
\end{equation}

The {\it reduced} density matrix for the system oscillator 
is introduced as 
\begin{eqnarray}
(\rho_S)_{mn}(t) 
&\equiv& {\rm Tr} \{ P_{mn}  \rho(t) \}
= {\rm Tr} \{ P_{mn}
e^{-i {\cal H}_0 t/\hbar} \tilde{\rho}(t) e^{i {\cal H}_0 t/\hbar}  \},
\\
P_{mn} &\equiv& |n \rangle \langle m| \otimes 1_B,
\\
\rho(0) &=& 
\rho_S \otimes \rho_B
=
\sum_{k,l} (\rho_S)_{kl} |k \rangle \langle l| 
\otimes e^{-\beta {\cal H}_B}/Z_B,
\\
Z_B &=& {\rm Tr}_B \{ e^{-\beta {\cal H}_B} \},
\end{eqnarray}
where the initial state is assumed to be a direct product 
state of $\rho_S$ and 
$\rho_B = e^{-\beta {\cal H}_B}/Z_B$, i.e.,
we have assumed that the bath is in thermal equilibrium.
Here $|k \rangle$ 
is the vibrational eigenstate for the system Hamiltonian ${\cal H}_S$,
i.e., ${\cal H}_S | k \rangle =E_k | k \rangle$.
If we assume that ${\cal H}_f$ is small, we can calculate $|k \rangle$
and $E_k$ using the time-independent perturbation theory as shown 
in Appendix \ref{app:system}.


We note that
\begin{equation}
(\rho_S)_{mn}(t)={\rm Tr} \{ P_{mn}
e^{-i {\cal H}_0 t/\hbar} \tilde{\rho}(t) e^{i {\cal H}_0 t/\hbar}  \}
=
{\rm Tr}_B \{ \tilde{\rho}_{mn}(t)  \} e^{-i \omega_{mn}t}.
\end{equation}
The lowest (second) order result for the density matrix is 
\begin{eqnarray}
(\rho_S)_{mn}(t)
&\simeq&
(\rho_S)^{(0)}_{mn}(t)
+(\rho_S)^{(1)}_{mn}(t)
+(\rho_S)^{(2)}_{mn}(t)
+\cdots,
\label{eq:density}
\\
(\rho_S)^{(0)}_{mn}(t)
&=&
{\rm Tr}_B \{ \tilde{\rho}_{mn}(0)  \} e^{-i \omega_{mn}t}
=(\rho_S)_{mn}e^{-i \omega_{mn}t},
\\
(\rho_S)^{(1)}_{mn}(t)
&=&
\frac{1}{i \hbar}
\int^{t}_0 dt' 
{\rm Tr}_B \{ \langle m| [\tilde{\cal V}(t'),\rho(0)] | n \rangle 
\} 
e^{-i \omega_{mn}t}
\nonumber
\\
&=&
\frac{1}{i \hbar}
\int^{t}_0 dt'
\sum_k
\left \{ 
\langle \tilde{\cal V}_{mk}(t') \rangle e^{i \omega_{mk}t'} (\rho_S)_{kn}
-
\langle \tilde{\cal V}_{kn}(t') \rangle e^{i \omega_{kn}t'} (\rho_S)_{mk}
\right \}
e^{-i \omega_{mn}t},
\\
(\rho_S)^{(2)}_{mn}(t)
&=&
\frac{1}{(i \hbar)^2}
\int^{t}_0 dt' 
\int^{t'}_0 dt'' 
{\rm Tr}_B \{ \langle m| [\tilde{\cal V}(t'),[\tilde{\cal V}(t''),\rho(0)]] | n \rangle 
\} 
e^{-i \omega_{mn}t}
\nonumber
\\
&=&
\frac{1}{(i \hbar)^2} \int_0^t dt' \int_0^{t'} dt''  
\sum_{k,l}
\left \{
\langle \tilde{\cal V}_{mk}(t') \tilde{\cal V}_{kl}(t'') \rangle 
(\rho_S)_{ln} e^{i (\omega_{mk}t'+\omega_{kl}t'')} 
\right \}
e^{-i \omega_{mn}t}
\nonumber
\\
&+&
\frac{1}{(i \hbar)^2} \int_0^t dt' \int_0^{t'} dt''  
\sum_{k,l}
\left \{
\langle \tilde{\cal V}_{kl}(t'') \tilde{\cal V}_{ln}(t') \rangle 
(\rho_S)_{mk} e^{i (\omega_{kl}t''+\omega_{ln}t')} 
\right \}
e^{-i \omega_{mn}t}
\nonumber
\\
&-&
\frac{1}{(i \hbar)^2} \int_0^t dt' \int_0^{t'} dt''  
\sum_{k,l}
\left \{
\langle \tilde{\cal V}_{ln}(t'') \tilde{\cal V}_{mk}(t') \rangle 
(\rho_S)_{kl} e^{i (\omega_{mk}t'+\omega_{ln}t'')} 
\right \}
e^{-i \omega_{mn}t}
\nonumber
\\
&-&
\frac{1}{(i \hbar)^2} \int_0^t dt' \int_0^{t'} dt''  
\sum_{k,l}
\left \{
\langle \tilde{\cal V}_{ln}(t') \tilde{\cal V}_{mk}(t'') \rangle 
(\rho_S)_{kl} e^{i (\omega_{mk}t''+\omega_{ln}t')} 
\right \}
e^{-i \omega_{mn}t}
\end{eqnarray}
where 
\begin{eqnarray}
\langle \tilde{\cal V}_{kl} (t) \tilde{\cal V}_{mn}(t') \rangle
&\equiv&
{\rm Tr}_B \{ \rho_B
\tilde{\cal V}_{kl}(t) \tilde{\cal V}_{mn}(t') \},
\\
\tilde{\cal V}_{kl}(t)
&=& \langle k | \tilde{\cal V}(t)| l \rangle
= \langle k | e^{i {\cal H}_B t/\hbar} {\cal V} e^{-i {\cal H}_B t/\hbar} | l \rangle,
\\
\omega_{kl} &=& (E_{k}-E_{l})/\hbar.
\end{eqnarray}
Note that, in the above formulas, the time dependence 
is only induced by the bath Hamiltonian ${\cal H}_B$.

For the matrix elements of the interaction Hamiltonian ${\cal V}$,
we have 
\begin{eqnarray}
\langle \tilde{\cal V}_{kl}(t) \rangle
&=&
-(q_S)_{kl} \langle \delta{\cal F}(t) \rangle
+(q_S^2)_{kl} \langle \delta{\cal G}(t) \rangle,
\\
\langle \tilde{\cal V}_{kl}(t) \tilde{\cal V}_{mn}(t') 
\rangle
&=&
(q_S)_{kl} (q_S)_{mn} 
\langle \delta{\cal F}(t) \delta{\cal F}(t') \rangle
+(q_S^2)_{kl} (q_S^2)_{mn} 
\langle \delta{\cal G}(t) \delta{\cal G}(t') \rangle
\nonumber
\\
&&
-[(q_S)_{kl} (q_S^2)_{mn}+(q_S^2)_{kl} (q_S)_{mn}] 
\langle \delta{\cal F}(t) \delta{\cal G}(t') \rangle
\label{eq:matrix}
\end{eqnarray}
where the value of $(q_S)_{kl}$ and $(q_S^2)_{kl}$ are 
given in Eqs.~(\ref{eq:qs1})-(\ref{eq:qs2}) for the case 
of a cubic oscillator.
Since $\langle \delta {\cal F} \rangle=0$ and 
$\langle \delta {\cal G} \rangle=0$,
we have $\langle \tilde{\cal V}_{kl}(t) \rangle=0$ and 
$(\rho_S)^{(1)}_{mn}(t)=0$.

\subsection{VER formula}

We first calculate the diagonal elements of the density matrix 
$(\rho_S)_{ii}(t)$ ($i=0,1$) by assuming that the initial state is 
the first {\it vibrationally} excited state: 
$\rho_S= |1 \rangle \langle 1|$. This is a typical situation 
for VER though VER from highly excited states can be 
considered \cite{VFVAMJ04}.
The density matrix $(\rho_S)_{00}(t)$
is written as 
\begin{eqnarray}
(\rho_S)_{00}(t)
&\simeq& 
\frac{2}{\hbar^2}
\int_0^t dt' \int_0^{t'} dt''  
{\rm Re} \left \{
\langle \tilde{\cal V}_{10}(t') \tilde{\cal V}_{01}(t'') \rangle 
e^{i \tilde{\omega}_S (t'-t'')} 
\right \}
\end{eqnarray}
where $\tilde{\omega}_S$ is the anharmonicity-corrected 
system frequency given by Eq.~(\ref{eq:freq}).

From Eq.~(\ref{eq:matrix}), we have
\begin{eqnarray}
(\rho_S)_{00}(t)
&\simeq& 
\frac{2}{\hbar^2}
(q_S)^2_{10}
\int_0^t dt' \int_0^{t'} dt''  
{\rm Re} \left \{
\langle \delta {\cal F}(t'-t'') \delta {\cal F}(0) \rangle 
e^{i \tilde{\omega}_S (t'-t'')} 
\right \}
\nonumber
\\
&&
+
\frac{2}{\hbar^2}
(q^2_S)^2_{10}
\int_0^t dt' \int_0^{t'} dt''  
{\rm Re} \left \{
\langle \delta {\cal G}(t'-t'') \delta {\cal G}(0) \rangle 
e^{i \tilde{\omega}_S (t'-t'')} 
\right \}
\nonumber
\\
&&
-\frac{4}{\hbar^2}
(q_S)_{10}(q^2_S)_{10}
\int_0^t dt' \int_0^{t'} dt''  
{\rm Re} \left \{
\langle \delta {\cal F}(t'-t'') \delta {\cal G}(0) \rangle 
e^{i \tilde{\omega}_S (t'-t'')} 
\right \}.
\label{eq:VER0}
\end{eqnarray}
Using the explicit expressions for the correlation functions \cite{FBS05a},
the final VER formula is obtained as
\begin{eqnarray}
(\rho_S)_{00}(t)
&\simeq&
\frac{2}{\hbar^2}
\sum_{\alpha,\beta} 
\left[
C_{--}^{\alpha \beta}  
u_t(\tilde{\omega}_S-\omega_{\alpha}-\omega_{\beta})
\nonumber
+C_{++}^{\alpha \beta}  
u_t(\tilde{\omega}_S+\omega_{\alpha}+\omega_{\beta})
+C_{+-}^{\alpha \beta}  
u_t(\tilde{\omega}_S-\omega_{\alpha}+\omega_{\beta})
\right]
\nonumber
\\
&&
+\frac{2}{\hbar^2}
\sum_{\alpha} 
\left[ 
C^{\alpha}_{-} u_t(\tilde{\omega}_S-\omega_{\alpha})
+
C^{\alpha}_{+} u_t(\tilde{\omega}_S+\omega_{\alpha})
\right]
\label{eq:VER}
\end{eqnarray}
where $u_t(\Omega)$ is defined as 
\begin{eqnarray}
u_t(\Omega)
&=&
\int_0^t dt' \int_0^{t'} dt''  
\cos \Omega (t'-t'') 
= \frac{1-\cos \Omega t}{\Omega^2},
\\
v_t(\Omega)
&=&
\int_0^t dt' \int_0^{t'} dt''  
\sin \Omega (t'-t'') 
= 
\frac{\Omega t-\sin \Omega t}{\Omega^2},
\end{eqnarray}
and $v_t(\Omega)$ is defined for later use.
The coefficients are defined in Appendix \ref{app:coef}.
Equation (\ref{eq:VER}) is our final formula for VER.

If we take the long time limit of this formula (which is equivalent to 
the Markov approximation), we obtain a formula for the VER rate
\begin{eqnarray}
k_{0 \leftarrow 1} 
&\equiv& 
\left. \frac{d}{dt}(\rho_S)_{00}(t) \right|_{t \rightarrow \infty}
\nonumber
\\
&=&
\frac{2 \pi}{\hbar^2}
\sum_{\alpha,\beta} 
\left[
C_{--}^{\alpha \beta}  
\delta(\tilde{\omega}_S-\omega_{\alpha}-\omega_{\beta})
+
C_{++}^{\alpha \beta}  
\delta(\tilde{\omega}_S+\omega_{\alpha}+\omega_{\beta})
+
C_{+-}^{\alpha \beta}  
\delta(\tilde{\omega}_S-\omega_{\alpha}+\omega_{\beta})
\right]
\nonumber
\\
&&
+\frac{2 \pi}{\hbar^2}
\sum_{\alpha} 
\left[ 
C^{\alpha}_{-} \delta(\tilde{\omega}_S-\omega_{\alpha})
+
C^{\alpha}_{+} \delta(\tilde{\omega}_S+\omega_{\alpha})
\right]
\label{eq:MF}
\end{eqnarray}
where we have used 
\begin{equation}
\left. \frac{d}{dt} u_t(\Omega) \right|_{t \rightarrow \infty} 
=\left. \frac{\sin \Omega t}{\Omega} \right|_{t \rightarrow \infty} 
= \pi \delta(\Omega).
\end{equation}
If $\bar{q}_S=q_S$ and $\tilde{\omega}_S=\omega_S$, i.e., 
$\langle {\cal F} \rangle = \langle {\cal G} \rangle = 0$ 
{\it and} $f=0$,
we recover the Maradudin-Fein formula \cite{FBS05a} from 
Eq.~(\ref{eq:MF}). It follows that Eq.~(\ref{eq:VER}) is a generalization of 
the Maradudin-Fein formula, which can describe the time development 
of a density matrix.

\subsection{Dephasing formula}

We calculate the off diagonal elements of the density matrix 
$(\rho_S)_{10}(t)$ by assuming
that the initial state is the superposition state 
between $|0 \rangle$ and  $|1 \rangle$: 
$\rho_S=(1/2)(
|0 \rangle \langle 0|+
|0 \rangle \langle 1|+
|1 \rangle \langle 0|+
|1 \rangle \langle 1|)
$ \cite{MO04}. 
That is, $(\rho_S)_{kl}=1/2$ for all $k$ and $l$.
This is a simplified situation to consider dephasing in a two-level system.
We have 
\begin{eqnarray}
(\rho_S)_{10}(t)
&=& (\rho_S)^{(0)}_{10}(t)+(\rho_S)^{(1)}_{10}(t)+(\rho_S)^{(2)}_{10}(t) + \cdots 
\nonumber
\\
&=&
\frac{1}{2}e^{-i \tilde{\omega}_S t}(1+ r^{(1)}(t)+r^{(2)}(t)+ \cdots).
\label{eq:dephasing}
\end{eqnarray}
By the definition of the interaction Hamiltonian (Appendix \ref{app:system}), 
we have $r^{(1)}(t)=0$.
The remaining term $r^{(2)}(t)$ is decomposed as
\begin{eqnarray}
r^{(2)}(t)
&=& -r^{(2)}_{FF}(t)-r^{(2)}_{GG}(t)-r^{(2)}_{FG}(t),
\\
r^{(2)}_{FF}(t)
&=&
\frac{2}{\hbar^2}
 \int_0^t dt' \int_0^{t'} dt''  
 {\rm Re} \{ \langle {\cal V}_{10}(t'){\cal V}_{01}(t'') \rangle \}
[e^{i \tilde{\omega}_S(t'-t'')}-e^{i \tilde{\omega}_S(t'+t'')}]
\nonumber
\\
&=&
\frac{2}{\hbar^2} 
(q_S)_{10}^2
 \int_0^t dt' \int_0^{t'} dt''  
 {\rm Re} \{ \langle \delta {\cal F}(t') \delta {\cal F}(t'') \rangle \}
[e^{i \tilde{\omega}_S(t'-t'')}-e^{i \tilde{\omega}_S(t'+t'')}]
\nonumber
\\
&+&
\frac{2}{\hbar^2} 
(q^2_S)_{10}^2
 \int_0^t dt' \int_0^{t'} dt''  
 {\rm Re} \{ \langle \delta {\cal G}(t') \delta {\cal G}(t'') \rangle \}
[e^{i \tilde{\omega}_S(t'-t'')}-e^{i \tilde{\omega}_S(t'+t'')}]
\nonumber
\\
&-&
\frac{4}{\hbar^2} 
(q_S)_{10} (q^2_S)_{10}
 \int_0^t dt' \int_0^{t'} dt''  
 {\rm Re} \{ \langle \delta {\cal F}(t') \delta {\cal G}(t'') \rangle \}
[e^{i \tilde{\omega}_S(t'-t'')}-e^{i \tilde{\omega}_S(t'+t'')}],
\\
r^{(2)}_{GG}(t)
&=&
\frac{1}{\hbar^2} \int_0^t dt' \int_0^{t'} dt''  
\{ 
\langle 
[{\cal V}_{11}(t')- {\cal V}_{00}(t')] {\cal V}_{11}(t'') 
\rangle 
+
\langle
{\cal V}_{00}(t'') [{\cal V}_{00}(t')- {\cal V}_{11}(t')]
\rangle  
\}
\nonumber
\\
&=&
\frac{1}{\hbar^2} [(q_S)_{11}-(q_S)_{00}]^2
\int_0^t dt' \int_0^{t'} dt''  
{\rm Re}
\langle 
\delta {\cal F}(t') \delta {\cal F}(t'') 
\rangle 
\nonumber
\\
&+&
\frac{1}{\hbar^2} [(q^2_S)_{11}-(q^2_S)_{00}]^2
\int_0^t dt' \int_0^{t'} dt''  
{\rm Re}
\langle 
\delta {\cal G}(t') \delta {\cal G}(t'') 
\rangle 
\nonumber
\\
&-&
\frac{2}{\hbar^2} [(q_S)_{11}-(q_S)_{00}] [(q^2_S)_{11}-(q^2_S)_{00}]
\int_0^t dt' \int_0^{t'} dt''  
{\rm Re}
\langle 
\delta {\cal F}(t') \delta {\cal G}(t'') 
\rangle 
\nonumber
\\
&+&
\frac{i}{\hbar^2} [(q_S)^2_{11}-(q_S)^2_{00}]
\int_0^t dt' \int_0^{t'} dt''  
{\rm Im}
\langle 
\delta {\cal F}(t') \delta {\cal F}(t'') 
\rangle 
\nonumber
\\
&+&
\frac{i}{\hbar^2} [(q^2_S)^2_{11}-(q^2_S)^2_{00}]
\int_0^t dt' \int_0^{t'} dt''  
{\rm Im}
\langle 
\delta {\cal G}(t') \delta {\cal G}(t'') 
\rangle 
\nonumber
\\
&-&
\frac{2 i}{\hbar^2} [(q_S)_{11} (q^2_S)_{11} - (q^2_S)_{00}(q^2_S)_{00}]
\int_0^t dt' \int_0^{t'} dt''  
{\rm Im}
\langle 
\delta {\cal F}(t') \delta {\cal G}(t'') 
\rangle, 
\label{eq:gg}
\\
r^{(2)}_{FG}(t)
&=&
\frac{2i}{(i \hbar)^2} \int_0^t dt' \int_0^{t'} dt''  
{\rm Im} 
\{ 
\langle [{\cal V}_{11}(t')-{\cal V}_{00}(t')] {\cal V}_{10}(t'') \rangle 
\}
[e^{i \tilde{\omega}_S t'}-e^{i \tilde{\omega}_S t''}]
\nonumber
\\
&=&
\frac{2i}{(i \hbar)^2} 
[(q_S)_{11}-(q_S)_{00}](q_S)_{10}
\int_0^t dt' \int_0^{t'} dt''  
{\rm Im} 
\{ 
\langle \delta {\cal F}(t') \delta {\cal F}(t'') \rangle 
\}
[e^{i \tilde{\omega}_S t'}-e^{i \tilde{\omega}_S t''}]
\nonumber
\\
&+&
\frac{2i}{(i \hbar)^2} 
[(q_S^2)_{11}-(q_S^2)_{00}] (q_S^2)_{10}
\int_0^t dt' \int_0^{t'} dt''  
{\rm Im} 
\{ 
\langle \delta {\cal G}(t') \delta {\cal G}(t'') \rangle 
\}
[e^{i \tilde{\omega}_S t'}-e^{i \tilde{\omega}_S t''}]
\nonumber
\\
&-&
\frac{2i}{(i \hbar)^2} 
\{ 
[(q_S)_{11}-(q_S)_{00}]
(q_S^2)_{10}
+
[(q_S^2)_{11}-(q_S^2)_{00}]
(q_S)_{10}
\}
\nonumber
\\
&&
\times
\int_0^t dt' \int_0^{t'} dt''  
{\rm Im} 
\{ 
\langle \delta {\cal F}(t') \delta {\cal G}(t'') \rangle 
\}
[e^{i \tilde{\omega}_S t'}-e^{i \tilde{\omega}_S t''}]
\end{eqnarray}
where the subscript denotes that, e.g., 
for $r^{(2)}_{FF}(t)$, the dominant contribution comes 
from $\langle \delta {\cal F}(t) \delta {\cal F}(0) \rangle$ 
when 
the effects of the bath and the system anharmonicity are both weak.

After similar calculations as done for the VER formula above, 
we obtain
\begin{eqnarray}
r^{(2)}_{FF}(t)
&=&
\frac{2}{\hbar^2}
\sum_{\alpha,\beta} 
\left[
(C_{--}^{\alpha \beta}  +C_{++}^{\alpha \beta})
f_t (\omega_{\alpha}+\omega_{\beta}) 
+C_{+-}^{\alpha \beta}  
f_t(\omega_{\alpha}-\omega_{\beta}) 
+
(C^{\alpha}_{-} + C^{\alpha}_{+}) f_t(\omega_{\alpha})
\right],
\label{eq:FF}
\\
r^{(2)}_{GG}(t)
&=&
\frac{1}{\hbar^2}
\sum_{\alpha,\beta}
\left[
(D_{--}^{R\alpha \beta}+D_{++}^{R\alpha \beta})
u_t(\omega_{\alpha}+\omega_{\beta})
+
D_{+-}^{R\alpha \beta}
u_t(\omega_{\alpha}-\omega_{\beta})
+
(D_{-}^{R \alpha}+D_{+}^{R\alpha})
u_t(\omega_{\alpha})
\right]
\nonumber
\\
&&
- \frac{i}{\hbar^2}
\sum_{\alpha,\beta}
\left[
(D_{--}^{I\alpha \beta}-D_{++}^{I\alpha \beta})
v_t(\omega_{\alpha}+\omega_{\beta})
+
D_{+-}^{I\alpha \beta}
v_t(\omega_{\alpha}-\omega_{\beta})
+(D^{I \alpha}_{-}-D^{I \alpha}_{+})
v_t(\omega_{\alpha})
\right],
\label{eq:GG}
\\
r^{(2)}_{FG}(t)
&=&
\frac{2}{\hbar^2}
\sum_{\alpha,\beta}
\left[
(E_{--}^{\alpha \beta}-E_{++}^{\alpha \beta})
g_t(\omega_{\alpha}+\omega_{\beta})
+E_{+-}^{\alpha \beta}
g_t(\omega_{\alpha}-\omega_{\beta})
+(E^{\alpha}_{-}-E^{\alpha}_{+})
g_t(\omega_{\alpha})
\right]
\label{eq:FG}
\end{eqnarray}
where 
\begin{eqnarray}
f_t(\Omega)
&=&
\int_0^t dt' \int_0^{t'} dt''  
\cos \Omega (t'-t'') 
[e^{i \tilde{\omega}_S(t'-t'')} -e^{i \tilde{\omega}_S(t'+t'')}]
\nonumber
\\
&=& 
\frac{1}{2}
\left[
\frac{1}{\tilde{\omega}_S-\Omega}+\frac{1}{\tilde{\omega}_S+\Omega}
\right]
\left\{
\frac{1-e^{i(\tilde{\omega}_S+\Omega)t}}{\tilde{\omega}_S+\Omega}
+\frac{1-e^{i(\tilde{\omega}_S-\Omega)t}}{\tilde{\omega}_S-\Omega}
+\frac{2i \tilde{\omega}_S t -1+e^{2 i\tilde{\omega}_S t}}{2\tilde{\omega}_S}
\right\},
\\
g_t(\Omega)
&=&
i \int_0^t dt' \int_0^{t'} dt''  
\sin \Omega (t'-t'') 
[e^{i \tilde{\omega}_S t'} -e^{i \tilde{\omega}_S t''}]
\nonumber
\\
&=&
\frac{
\tilde{\omega}_S (1+e^{i \tilde{\omega}_S t}) (1-\cos \Omega t )
-i \Omega (1-e^{i \tilde{\omega}_S t}) \sin \Omega t } 
{\Omega (\tilde{\omega}_S^2-\Omega^2)},
\end{eqnarray}
and the coefficients are defined in Appendix \ref{app:coef}.
Equation (\ref{eq:dephasing}) and Eqs.~(\ref{eq:FF})-(\ref{eq:FG}) 
constitute the dephasing formula.

Dephasing properties are characterized by the decaying behavior of 
this off diagonal density matrix.
Incidentally, as an alternative approach, 
one might use the von Neuman entropy or linear entropy for the 
reduced system as an indicator of dephasing \cite{FMT03}.

\subsection{Frequency autocorrelation function}

Using the time-independent perturbation theory 
for the interaction, we obtain the fluctuation of the 
system frequency as 
\begin{equation}
\delta \omega(t)= \frac{{\cal V}_{11}(t)-{\cal V}_{00}(t)}{\hbar}
=-\frac{(q_S)_{11}-(q_S)_{00}}{\hbar} \delta {\cal F}(t)
+\frac{(q_S^2)_{11}-(q_S^2)_{00}}{\hbar} \delta {\cal G}(t).
\end{equation}
Hence we have 
\begin{eqnarray}
{\rm Re}\langle \delta \omega(t')\delta \omega(t'') \rangle
&=& 
\frac{1}{\hbar^2}
\left(
[(q_S)_{11}-(q_S)_{00}]^2
{\rm Re} \langle \delta {\cal F}(t')\delta {\cal F}(t'') \rangle
+[(q_S^2)_{11}-(q_S^2)_{00}]^2
{\rm Re} \langle \delta {\cal G}(t')\delta {\cal G}(t'') \rangle
\right.
\nonumber
\\
&&
\left.
-2 [(q_S)_{11}-(q_S)_{00}]
[(q_S^2)_{11}-(q_S^2)_{00}]
{\rm Re} \langle \delta {\cal F}(t')\delta {\cal G}(t'') \rangle
\right).
\end{eqnarray}
This turns out to be the second derivative of 
${\rm Re} \{ r^{(2)}_{GG}(t) \}$, i.e.,
\begin{equation}
C(t) \equiv {\rm Re}\langle \delta \omega(t)\delta \omega(0) \rangle
=
\frac{d^2}{dt^2} {\rm Re} \{ r^{(2)}_{GG}(t) \}.
\end{equation}
From this correlation function, 
we can define a {\it pure dephasing time} $T_2^*$ as
\begin{eqnarray}
\frac{1}{T_2^*}
=
\int_0^{\infty} C(t) dt. 
\end{eqnarray}
Note that this is different from a correlation time defined by
\begin{eqnarray}
\tau_c
=
\frac{1}{C(0)} \int_0^{\infty} C(t) dt 
\end{eqnarray}
which leads to the relation $1/T_2^*=C(0) \tau_c$.
Note that $T_2^*$ and $\tau_c$ are inversely related.


\section{Application: NMA in heavy water}
\label{sec:NMA}

\subsection{NMA-D in heavy water}

We now apply our formulas to VER and dephasing problems 
of $N$-methylacetamide (NMA) in heavy water.
In many theoretical and experimental studies,
this molecule CH$_3$-NH-CO-CH$_3$ is taken to be 
a model ``minimal'' peptide system 
because it contains a peptide bond (-NH-CO-).
For example, Gerber and coworkers calculated the 
vibrational frequencies for this molecule
using the vibrational self-consistent field (VSCF) method \cite{GCG02}. 
Nguyen and Stock worked to characterize VER in
this molecule using a quasi-classical method \cite{NS03}.
Skinner and coworkers investigated the dephasing properties of 
the amide I mode using their correlation method combined with ab initio (DFT) 
calculations \cite{SCS04}.
Employing 2D-IR spectroscopy, Zanni {\it et al}.~\cite{ZAH01}
measured $T_1$ and $T_2^*$ for the amide I mode in this molecule,
which were reported to be $T_1 \simeq 0.45$ ps and $T_2^* \simeq 1.12$ ps,
whereas Woutersen {\it et al}. \cite{WPHMKS02} obtained  $T_2^* \simeq 0.8$ ps.

In our study, we deuterate the system to NMAD/D$_2$O so that 
the amide I mode,
localized around the CO bond,
can be clearly recognized as a single peak in the spectrum. 
In the following numerical calculations 
based on the CHARMM force field \cite{CHARMM}, 
its frequency is $\sim$ 1690 cm$^{-1}$ which fluctuates depending on the structure 
(see Fig.~\ref{fig:shift}),
whereas the experimental and DFT values are 1717 cm$^{-1}$ and 
1738 cm$^{-1}$, respectively \cite{SCS04}.

\subsection{Procedure}

We have applied the following general procedure.
(1) Run an equilibrium simulation. 
(2) Sample several trajectories during the run.
(3) Delete atoms of each configuration except the ``active'' region 
around the system oscillator.
We introduce a cut off radius $R_c$ around a certain 
atom within the active site.
(4) Calculate instantaneous normal modes (INMs) \cite{Keyes} for each ``reduced'' configuration,
ignoring all the imaginary frequencies \cite{MO04}.
(5) Calculate anharmonic coupling elements with the finite difference method \cite{FBS05a} 
using the obtained INMs.
(6) Insert the results in the VER formula [Eq.~(\ref{eq:VER})] and dephasing formula 
[Eq.~(\ref{eq:dephasing}) with Eqs.~(\ref{eq:FF})-(\ref{eq:FG})]. 
(7) Ensemble average the resultant density matrix, and 
estimate the VER and dephasing rates (times), if possible. 

\begin{figure}[htbp]
\hfill
\begin{center}
\includegraphics[scale=1.2]{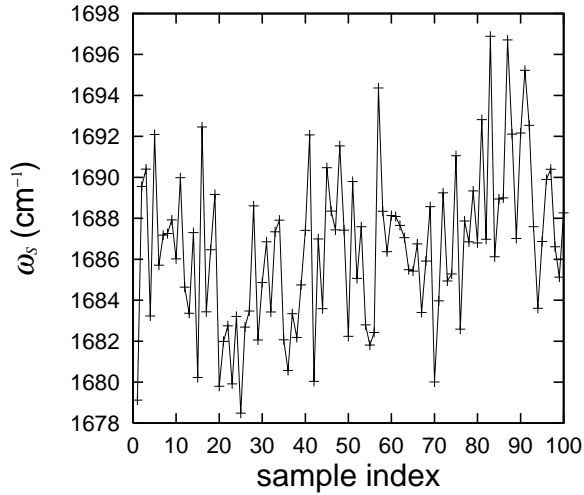}
\end{center}
\caption{
Instantaneous normal mode frequency of the system
${\omega}_S$ 
for 100 different sample trajectories at 300K where
the cut off radius is $R_c= 10$ \AA.
}
\label{fig:shift}
\end{figure}

This procedure seems to be straighforward.
However, when applied to real systems like peptides or proteins,
we need to carefully treat the the effects of the bath.
In Fig.~\ref{fig:shift}, we plot the system frequency ${\omega}_S$ 
for 100 sample trajectories from an equilibrium run at 300K.
We see that the amide I mode frequency changes depending on 
the structure; the deviation can amount to 1\%. 


{\it Furthermore} the frequency is renormalized according to Eqs.~(\ref{eq:omegas}) and (\ref{eq:freq}),
and such an effect can be anomalously large if we include all the contribution from 
low frequency components. Hence we need to introduce a cutoff frequency $\omega_c$,
below which the contribution is neglected.
This is physically sound, because we are dealing with time-dependent phenomena,
and such low frequency components correspond to longer time behavior. However,
we are now interested in rather short time dynamics, so such contributions 
should not play a role. In fact, the final result of VER does not depend 
much on the choice of $\omega_c$, whereas that of dephasing does.
We need to admit that for now this is just a remedy.
We discuss how to improve this situation later.


\subsection{VER properties of NMA-D}

\begin{figure}[htbp]
\hfill
\begin{center}
\includegraphics[scale=1.2]{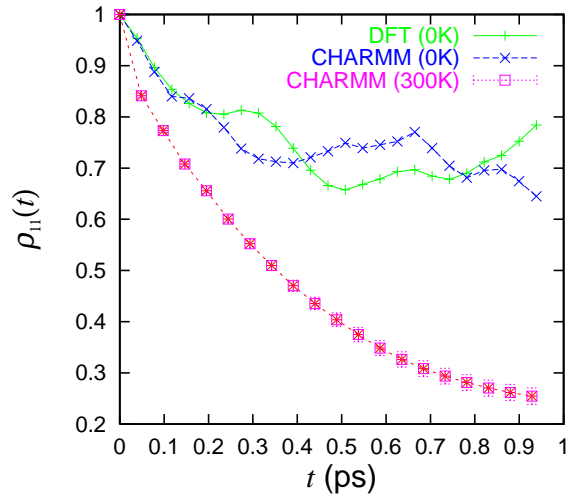}
\end{center}
\caption{
Time evolution of the excited density matrix 
for NMA-D in vacuum (DFT and CHARMM) at 0K, and for 
NMA-D in heavy water (CHARMM) at 300K.
The level of DFT is B3LYP/6-31+G(d).
}
\label{fig:VER1}
\end{figure}

First we consider the VER properties of the amide I mode as shown in
Fig.~\ref{fig:VER1}.
We use the following relation 
\begin{equation}
\rho_{11}(t)=1 - \rho_{00}(t)=\exp[-s(t)]
\end{equation}
and hypothesize that $s(t) \simeq \rho_{00}(t)$, which is definitely 
true  when $\rho_{00}(t) \ll 1$, and might be justified using 
the cumulant expansion technique \cite{Breuer,IMM03}.

We calculated the density matrix for the following three cases: 
(a) NMA-D in heavy water with CHARMM force field at 300K,
(b) NMA-D in vacuum with CHARMM force field at 0K,
and 
(c) NMA-D in vacuum with DFT force field at 0K.
Here we have used $R_c=10$ \AA \, and $\omega_c=10$ cm$^{-1}$ for case (a).
The result for VER does not depend sensitively on these parameters.
For cases (b) and (c), we must take a special care:
It is known that the low frequency components cause serious 
problems for vibrational frequency calculations \cite{Yagi},
so we need to eliminate the low frequency components.
In this work, we exclude several normal modes if their frequency 
is less than 300 cm$^{-1}$. See Table \ref{table:NMAfreq}.

\begin{table}[htbp]
\caption{
Normal mode frequencies (in cm$^{-1}$) 
for ab initio (left) and CHARMM (right) NMA.
The level of the ab initio calculation is B3LYP/6-31+G(d).
}
\hfill
\begin{center}
\begin{tabular}{c|c|c}
\hline
Mode index $\alpha$   & $\omega_{\alpha}$ (ab initio) & $\omega_{\alpha}$ (CHARMM) \\
\hline \hline
1 &      31.5 	       &   64.1    \\ \hline
2 &      71.6	       &   88.9    \\ \hline
3 &     170.2	       &  192.3    \\ \hline
4 &     259.6	       &  269.5    \\ \hline
5 &     282.3	       &  426.7    \\ \hline
6 &     421.9	       &  536.3    \\ \hline
7 &     619.1	       &  575.9    \\ \hline
8 &     619.9	       &  741.9    \\ \hline
9 &     868.7	       &  757.0    \\ \hline
10 &    946.1	       &  854.4    \\ \hline
11 &   1012.4	       &  964.3    \\ \hline
12 &   1066.1	       & 1055.9    \\ \hline
13 &   1144.5	       & 1075.7    \\ \hline
14 &   1158.9	       & 1088.5    \\ \hline
15 &   1207.6	       & 1123.5    \\ \hline
\end{tabular}
\hspace{1cm}
\begin{tabular}{c|c|c}
\hline
Mode index $\alpha$   & $\omega_{\alpha}$ (ab initio) & $\omega_{\alpha}$ (CHARMM) \\
\hline \hline
16 &   1417.9	       &     1380.2    \\ \hline
17 &   1436.1	       &     1408.7    \\ \hline
18 &   1483.6	       &     1415.4    \\ \hline
19 &   1495.1	       &     1418.5    \\ \hline
20 &   1499.4	       &     1425.7    \\ \hline
21 &   1516.7	       &     1444.7    \\ \hline
22 &   1535.9	       &     1563.1    \\ \hline
23 &   1745.9	       &     1678.1    \\ \hline
24 &   2671.1	       &     2445.0    \\ \hline
25 &   3058.5	       &     2852.8    \\ \hline
26 &   3058.9	       &     2914.3    \\ \hline
27 &   3116.5	       &     2914.8    \\ \hline
28 &   3130.9	       &     2917.2    \\ \hline
29 &   3135.9	       &     2975.3    \\ \hline
30 &   3148.9	       &     2975.5    \\ \hline
\end{tabular}
\end{center}
\label{table:NMAfreq}
\end{table}

By using a fitting form $s(t)=t/T_1$,
where $T_1$ is the VER time,
we estimate that 
$T_1 \simeq 0.5$ ps at 300K and 0.6 ps at 0K from the initial decay. 
The former estimate is rather similar to the experimental value $T_1 \simeq 0.45$ ps \cite{ZAH01},
whereas Nguyen-Stock's quasi-classical estimate is $T_1 \simeq 1.5$ ps \cite{NS03}.
Considering that the estimate at 300K is rather close to that at 0K,
we can conclude that quantum effects are important to describe VER for 
the amide I mode of NMA-D in heavy water.
However, the decay at the later stage becomes very slow at 0K as expected 
because there is no environment. 
(In the vacuum cases, we only use one minimized structure, 
thus there is no ensemble average, and the oscillatory behavior remains.)

The results are similar for 
NMA-D in vacuum with different force fields.
It is known that NMA with CHARMM force field is not well
characterized around the methyl groups \cite{GK92}, but
this fact does not affect the VER properties of the amide I mode.

\begin{figure}[htbp]
\hfill
\begin{center}
\includegraphics[scale=1.2]{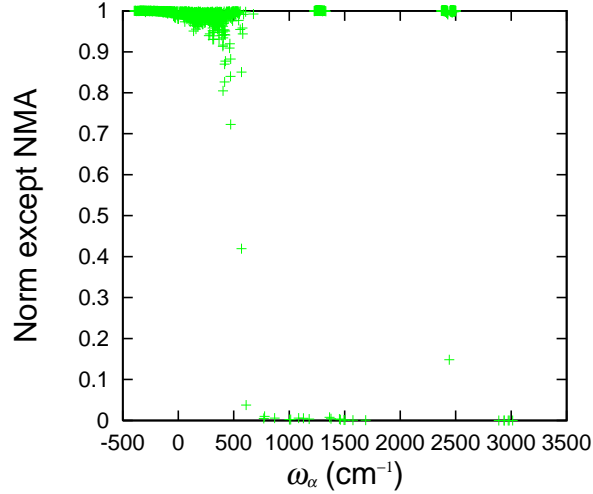}
\end{center}
\caption{
Norm of the eigenvectors (normal modes)
with the exception of the contribution from NMA-D, 
which is defined by
$\sum_{i \in {\rm Water}} (x_i^2 +y_i^2+z_i^2)$,
where $i$ comes from water degrees of freedom alone.
}
\label{fig:norm}
\end{figure}

We have analyzed the mechanism of VER in terms of the VER pathway.
In Table \ref{table:pathway}, we show 
several mode combinations that contribute most to $s(t)$ for NMA-D 
in heavy water at 300K.
These eigenvectors (normal modes) are well localized 
around NMA (Fig.~\ref{fig:norm}), especially on the CO bond (Table \ref{table:norm}).
There is very little contribution from the surrounding water.
(This is expected from the previous result of Kidera and coworkers \cite{Kidera}.)
Similar ``resonant'' mode combinations can be found in the 
isolated NMA-D cases. See Tables \ref{table:pathway2} and \ref{table:pathway3}.
This means that the initial stage of VER of NMA-D in heavy water 
is dominated by intramolecular vibrational redistribution (IVR) 
localized near the peptide bond.
This result might explain why the amide I mode, 
in many peptide systems with differing environments, 
appears to have similar VER times \cite{MKFKAZ04}.
Note that this is the case for a {\it localized mode} such as the 
deuterated amide I mode. 
A {\it collective mode} can decay 
with a different VER pathway,
as shown by Austin's group \cite{XMHA00}.

\begin{table}[htbp]
\caption{
The most dominant VER pathways for the amide I mode of NMA-D in heavy water.
$\Delta \omega$ is defined by $|\omega_S-\omega_{\alpha}-\omega_{\beta}|$.
}
\hfill
\begin{center}
\begin{tabular}{c|c|c|c}
\hline \hline
Mode combination  ($\alpha, \beta$) &  frequency (cm$^{-1}$) &  Contribution to $s(t)$ & 
            $\Delta \omega$ (cm$^{-1}$) \\ \hline \hline           
       1143   +     1143 & 778.6   +    778.6  & 0.04 &       125.7    \\ \hline
       1147   +     1134 & 1085.3  +    570.0  & 0.04 &       27.6    \\ \hline
       1147   +     1135 & 1085.3  +    570.9  & 0.01 &       26.6      \\ \hline
       1147   +     1136 & 1085.3  +    578.8  & 0.02 &       18.8      \\ \hline
       1147   +     1137 & 1085.3  +    581.3  & 0.03 &       16.2      \\ \hline
       1147   +     1140 & 1085.3  +    612.1  & 0.46 &       14.5      \\ \hline
       1148   +     1132 & 1127.8  +    558.5  & 0.01 &       3.4      \\ \hline
       1148   +     1134 & 1127.8  +    570.0  & 0.11 &       14.9      \\ \hline
       1148   +     1135 & 1127.8  +    570.9  & 0.03 &       15.9      \\ \hline
\end{tabular}
\end{center}
\label{table:pathway}
\end{table}

\begin{table}[htbp]
\caption{
The most localized modes around the CO bond in NMA-D.
The norm is defined by $\sum_{i \in {\rm CO bond}} (x_i^2 +y_i^2+z_i^2)$.
The 1345th mode is the amide I mode.
}
\hfill
\begin{center}
\begin{tabular}{c|c|c}
\hline \hline
Mode index $\alpha$ &  frequency (cm$^{-1}$) & Contribution to norm  \\ \hline \hline           
        1134 & 570.0 & 0.14    \\ \hline
        1140 & 612.1 & 0.26    \\ \hline
        1142 & 771.2 & 0.35    \\ \hline
        1143 & 778.6 & 0.41    \\ \hline
        1146 & 1013.6 & 0.13    \\ \hline
        1147 & 1085.3 & 0.26    \\ \hline
        1148 & 1127.8 & 0.17    \\ \hline
        1340 & 1452.1 & 0.36    \\ \hline
        1345 & 1689.6 & 0.92    \\ \hline
\end{tabular}
\end{center}
\label{table:norm}
\end{table}

\begin{table}[htbp]
\caption{
The most dominant VER pathways for 
NMA-D in vacuum with the ab initio potential (B3LYP/6-31+G(d)).
}
\hfill
\begin{center}
\begin{tabular}{c|c|c|c}
\hline
Mode combination  ($\alpha, \beta$) &  frequency (cm$^{-1}$) &  Contribution to $s(t)$ &  
$\Delta \omega$ (cm$^{-1}$) \\ \hline \hline           
           9 +      9 & 868.7 + 868.7 & 0.13         &     2.7  \\ \hline
          13 +      8 & 1144.5+ 620.0 & 0.02         &     29.8 \\ \hline
\end{tabular}
\end{center}
\label{table:pathway2}
\end{table}

\begin{table}[htbp]
\caption{
The most dominant VER pathways for 
NMA-D in vacuum with the CHARMM force field.
}
\hfill
\begin{center}
\begin{tabular}{c|c|c|c}
\hline
Mode combination  ($\alpha, \beta$) &  frequency (cm$^{-1}$) & Contribution to $s(t)$ &  
 $\Delta \omega$ (cm$^{-1}$) \\ \hline \hline           
           8   +    8 & 741.9 +741.9 & 0.02  &  194.1   \\ \hline
          14   +    6 & 1088.5+536.3 & 0.08  &  53.1   \\ \hline
          14   +    8 & 1088.5+741.9 & 0.02  &  152.5   \\ \hline
          15   +    7 & 1123.5+575.9 & 0.08  &  21.6   \\ \hline
\end{tabular}
\end{center}
\label{table:pathway3}
\end{table}


\subsection{Dephasing properties of NMA-D}

We now consider the dephasing properties of the amide I mode.
The off-diagonal density matrix is written as 
\begin{equation}
\rho_{10}(t)
=\frac{1}{2} e^{- i \tilde{\omega}_S t} 
[1 - r^{(2)}_{FF}(t)-r^{(2)}_{GG}(t)-r^{(2)}_{FG}(t)]
\simeq 
\frac{1}{2} e^{- i \tilde{\omega}_S t - r^{(2)}_{FF}(t)-r^{(2)}_{GG}(t)-r^{(2)}_{FG}(t)}
\end{equation}
and we analyze each contribution to the density matrix seperately.

\begin{figure}[htbp]
\hfill
\begin{center}
\includegraphics[scale=1.2]{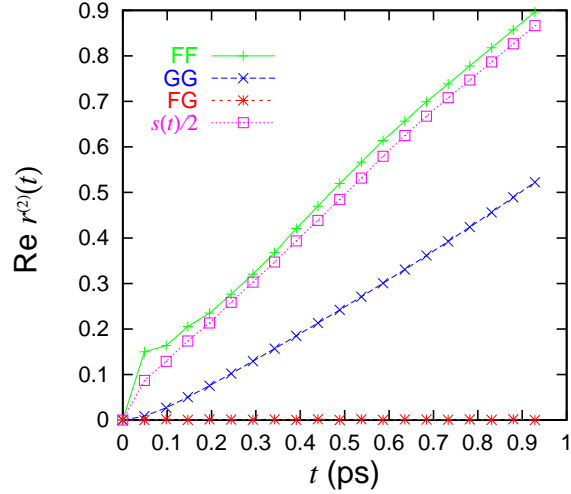}
\end{center}
\caption{
Dephasing properties of the amide I mode of NMA-D in heavy water.
}
\label{fig:deph}
\end{figure}

In Fig.~\ref{fig:deph}, we show the result with $R_c=10$ \AA \, and $\omega_c=10$ cm$^{-1}$.
We can see that the following 
relation holds
\begin{equation}
{\rm Re} \{ r^{(2)}_{FF}(t)\} \simeq s(t)/2 \simeq t/(2 T_1).
\end{equation}
If we {\it further} assume that ${\rm Re} \{ r^{(2)}_{GG}(t) \}\simeq t/T_2^*$ and 
${\rm Re} \{ r^{(2)}_{FG}(t) \} \simeq 0$,
we have 
\begin{equation}
\frac{1}{T_2}=\frac{1}{2T_1}+\frac{1}{T_2^*}.
\label{eq:T2}
\end{equation}
This is a standard expression connecting $T_1$ and $T_2$ \cite{Mchale}, 
and holds under the Markov approximation. 
We can see that ${\rm Re} \{ r^{(2)}_{FG}(t) \}\simeq 0$ holds,
but it is difficult to judge whether ${\rm Re} \{ r^{(2)}_{GG}(t) \} \simeq t/T_2^*$ holds or not.


\begin{figure}[htbp]
\hfill
\begin{center}
\begin{minipage}{.42\linewidth}
\includegraphics[scale=1.0]{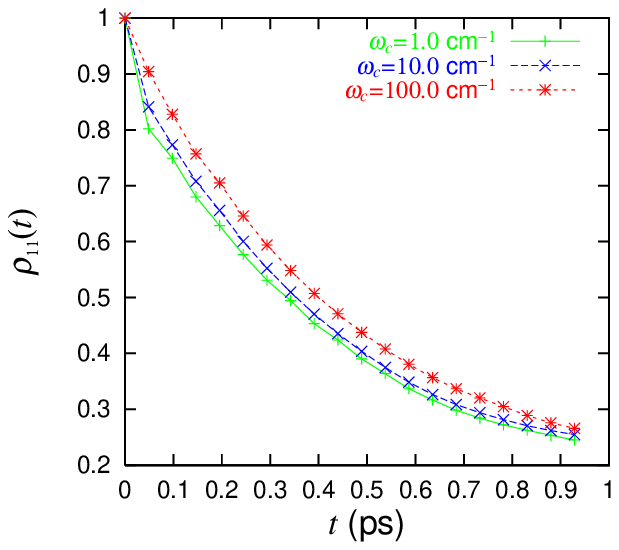}
\end{minipage}
\hspace{1cm}
\begin{minipage}{.42\linewidth}
\includegraphics[scale=1.0]{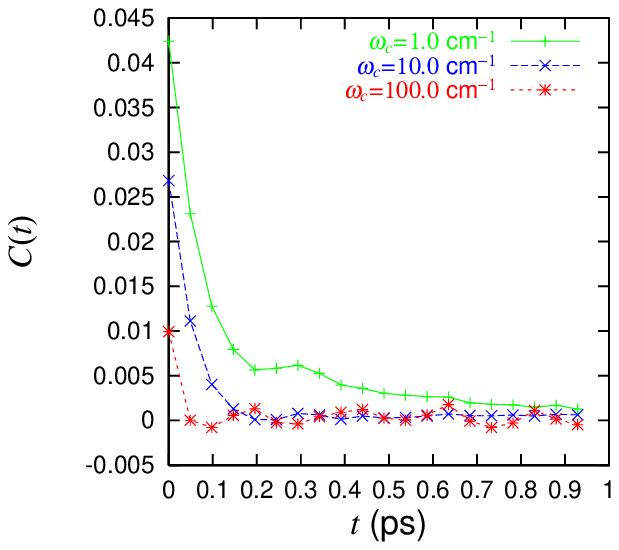}
\end{minipage}
\end{center}
\caption{
Left: VER (left) and frequency auto correlation 
calculations (right) at 300K with different 
cut-off frequencies $\omega_c$.
}
\label{fig:comp}
\end{figure}

There are more serious problems:
as mentioned by Mikami and Okazaki \cite{MO04},
the diagonal terms contribute most for dephasing, i.e.,
the second term in Eq.~(\ref{eq:GG}) is a dominant contribution for dephasing.
Furthermore, the coefficients are 
dominant factors, so this means that the low frequency (and thus delocalized) 
modes contribute most. 
In this paper, we have employed two cut-off parameters: $R_c$ and $\omega_c$.
If $R_c$ is large enough, it is fine, but the choice of $\omega_c$ can be 
arbitrary. Figure \ref{fig:comp} shows the dependence of the results 
on $\omega_c$. The VER results do not depend on the choice of $\omega_c$ 
because there is a resonant condition which should be met, but 
the dephasing results do.
We need to be cautious in the interpretation of our results for dephasing.
One way to get rid of this problem is to go back to 
the original expression Eq.~(\ref{eq:gg}) 
using the force and force-constant autocorrelation functions
$\langle \delta {\cal F}(t) \delta {\cal F}(0) \rangle$ and 
$\langle \delta {\cal G}(t) \delta {\cal G}(0) \rangle$.
Here $r^{(2)}_{GG}(t)$ is calculated as time integral of 
these correlation functions, which can be calculated using 
classical mechanics. This is in the same spirit as the 
quantum correction factor method \cite{SP01}, which is 
an approximation to quantum effects.
In this case, we only need to consider the 
zero frequency component, so the classical mechanics should work well
and quantum effects should be less important.

\subsection{Discussions}


We found that there are several resonant modes in 
NMA-D, which form the main VER pathways {\it within} the molecule.
Gerber and coworkers reported that the amide I mode in NMA 
is very weakly coupled to other modes \cite{GCG02}. 
We expect that this discrepancy results from 
(a) the use of only pair interactions between normal modes to reduce the computational cost,
(b) the level of the ab initio method: they
used MP2/DZP whereas we used B3LYP/6-31+G(d),
and 
(c) the criterion of the mode-mode coupling: 
their criterion is not directly related to VER.

It is important and interesting to clarify the nature of 
VER in the amide I mode in more detail. 
Note the importance of the system anharmonicity.
The effect of the system anharmonicity defined in Eq.~(\ref{eq:anhar}) 
is very weak for the CHARMM case:
$\varepsilon=10^{-5}$, but it is not for the ab initio case: 
$\varepsilon=10^{-2}$. According to Eq.~(\ref{eq:freq}), 
this anharmonicity shifts the system frequency by 0.6\%, which amounts 
to 10 cm$^{-1}$. The resonant condition changes compared 
to the case without anharmonicity. 
Of course, dephasing is also affected by this amount of anharmonicity. 
To address these issues, we must develop QM/MM type methods, which will be described elsewhere.
Another interesting system to investigate anharmonicity is 
a highly excited bond such as 
the highly excited CO bond in myoglobin \cite{VFVAMJ04}. 
 

It is important to assess our strategy: 
our perturbative expansion and cut-off strategy are 
approximate. It would be profitable and interesting to compare this strategy 
with others, including the time-dependent vibrational self-consistent field methods \cite{JG99,FYZSH},
the semiclassical method \cite{Geva}, and the path integral method \cite{Krilov}.
The application of our strategy to protein systems, including cytochrome c, 
will be described elsewhere \cite{Matt}.

\section{Summary}
\label{sec:summary}

In this paper,
we have derived formulas for VER and dephasing for an 
anharmonic (cubic) oscillator coupled to a harmonic bath 
through 3rd and 4th order coupling elements.
We employed time-dependent perturbation theory and did not
take the infinite time limit as is done in the derivation of the Maradudin-Fein formula.
Hence our formulas do not assume the Markov properties 
of the system, and can describe short time behavior 
that can be important for VER and dephasing properties 
of localized modes in peptides or proteins.
Our final results are the VER formula 
[Eq.~(\ref{eq:VER})] and dephasing formula 
[Eq.~(\ref{eq:dephasing}) with Eqs.~(\ref{eq:FF})-(\ref{eq:FG})]. 
As a test case, we have studied the amide I mode of 
$N$-methylacetamide in heavy water.
We found that the VER time is 0.5 ps at 300K,
which is in good accord with the experimental value,
and clarified that the VER mechanism is mainly 
localized around the peptide bond in NMA-D;
VER is dominated by IVR within the molecule.
We also investigated the dephasing properties of the 
amide I mode, and met with some problems.
We proposed a new method to overcome these problems 
using classical correlation function calculations.

\acknowledgments

We thank S.~Okazaki, T.~Mikami, K.~Yagi, T.~Miyadera, A.~Szabo, E.~Geva, G.~Krilov,
H.-P.~Breuer, S.~Maniscalco, 
F.~Romesberg and M.~Cremeens for useful discussions. We also thank the National Science 
Foundation (CHE-036551) and Boston University's Center for 
Computer Science for generous support to
our research.

\begin{appendix}

\section{System parameters for a cubic oscillator}
\label{app:system}

We assume that the system-bath interaction can be 
Taylor expanded using the bath coordinate $q_{\alpha}$,
and that the fluctuating force and 
the fluctuating force constant can be expressed as 
\begin{eqnarray}
\delta{\cal F} 
&=&
\sum_{\alpha,\beta}C_{S \alpha \beta} 
(q_{\alpha} q_{\beta} 
-\langle q_{\alpha} q_{\beta} \rangle),
\\
\delta{\cal G}
&=&
\sum_{\alpha,\beta}C_{SS \alpha \beta} 
(q_{\alpha} q_{\beta}
-\langle q_{\alpha} q_{\beta} \rangle)
+\sum_{\alpha}C_{SS \alpha} q_{\alpha}. 
\end{eqnarray}

In real molecular systems such as peptides or proteins, the coefficients 
in ${\cal V}$ and the anharmonicity parameter in ${\cal H}_f$ 
are calculated as  
\begin{eqnarray}
C_{S \alpha \beta}
 &=& 
-\frac{1}{2} \frac{\partial^3 V}{\partial q_S \partial q_{\alpha} \partial q_{\beta}},
\\
C_{SS \alpha}
 &=& 
\frac{1}{2} \frac{\partial^3 V}{\partial q_S^2 \partial q_{\alpha}},
\\
C_{SS \alpha \beta}
 &=& 
\frac{1}{4} \frac{\partial^4 V}{\partial q_S^2 \partial q_{\alpha} \partial q_{\beta}},
\\
f
&=&
\frac{\partial^3 V}{\partial q_S^3}
\end{eqnarray}
where $V$ represents a potential function for the system considered.
This potential function can be an empirical force field (CHARMM, Amber) or 
an ab initio potential calculated by any level of theory.

Assuming that the cubic anhamonicity $f$ in the system 
is small, we use the time-independent perturbation 
theory to calculate the eigen energies and vectors.
We quote from J.J.~Sakurai \cite{JJ}:
\begin{eqnarray}
E_n
&=& 
E^{(0)}_n+ V_{nn}+ \sum_{k \neq n} \frac{|V_{nk}|^2}{E^{(0)}_n-E^{(0)}_k},
\\
|n \rangle 
&=& |n^{(0)} \rangle 
+
\sum_{k \neq n} |k^{(0)} \rangle  
\frac{V_{kn}}{E^{(0)}_n-E^{(0)}_k}
\nonumber
\\
&&
+
\left(
\sum_{k \neq n} 
\sum_{l \neq n} 
|k^{(0)} \rangle  
\frac{V_{kl}V_{ln}}
{(E^{(0)}_n-E^{(0)}_k)(E^{(0)}_n-E^{(0)}_l)}
-\sum_{k \neq n} 
|k^{(0)} \rangle  
\frac{V_{nn}V_{kn}}
{(E^{(0)}_n-E^{(0)}_k)^2}
\right)
\end{eqnarray}
where $E^{(0)}_n=\hbar \bar{\omega}_S (n+1/2)$,
$|k^{(0)} \rangle$ is the
$k$-th eigenfunction of the harmonic oscillator,
and 
\begin{eqnarray}
V_{kn} 
&=& \frac{f}{6} 
\langle k^{(0)}| \bar{q}_S^3 |n^{(0)} \rangle
\nonumber
\\
&=& \hbar \bar{\omega}_S \varepsilon
\left[
\sqrt{n(n-1)(n-2)} \delta_{k,n-3}
+3n \sqrt{n} \delta_{k,n-1}
\right.
\nonumber
\\
&&
\left.
+3(n+1) \sqrt{n+1} \delta_{k,n+1}
+\sqrt{(n+1)(n+2)(n+3)} \delta_{k,n+3}
\right]
\end{eqnarray}
where 
\begin{equation}
\varepsilon=\frac{1}{\hbar \bar{\omega}_S} 
\frac{f}{6} \left( \frac{\hbar}{2 \bar{\omega}_S} \right)^{3/2} 
\label{eq:anhar}
\end{equation}
is a dimensionless paramater representing the strength of the anharmonicity of the system.
Note that $V_{kn}$ becomes nonzero only when $|k-n|=1$ or $|k-n|=3$.

We explicitly have 
\begin{eqnarray}
E_0
&=&
\frac{\hbar \bar{\omega}_S}{2}- 
\frac{|V_{01}|^2}{\hbar \bar{\omega}_S}
-\frac{|V_{03}|^2}{3 \hbar \bar{\omega}_S}
=
\frac{\hbar \bar{\omega}_S}{2}(1-22 \varepsilon^2),
\\
E_1
&=&
\frac{3 \hbar \bar{\omega}_S}{2}+ 
\frac{|V_{10}|^2}{\hbar \bar{\omega}_S}
-\frac{|V_{12}|^2}{\hbar \bar{\omega}_S}
-\frac{|V_{14}|^2}{3 \hbar \bar{\omega}_S}
=
\frac{\hbar \bar{\omega}_S}{2}(3-142 \varepsilon^2).
\end{eqnarray}
The anharmonicity-corrected frequency is 
\begin{equation}
\tilde{\omega}_S=\frac{E_1-E_0}{\hbar}
= 
\bar{\omega}_S (1-60 \varepsilon^2).
\label{eq:freq}
\end{equation}

Next we calculate the matrix elements for ${q}_S$ and ${q}_S^2$.
We write the eigenfunctions:
\begin{eqnarray}
|0 \rangle 
&=& |0^{(0)} \rangle 
+
\sum_{k=1,3} |k^{(0)} \rangle  
\frac{V_{k0}}{E^{(0)}_0-E^{(0)}_k}
+
\sum_{k,l \in S_0} 
|k^{(0)} \rangle  
\frac{V_{kl}V_{l0}}
{(E^{(0)}_0-E^{(0)}_k)(E^{(0)}_0-E^{(0)}_l)},
\\
|1 \rangle 
&=& |1^{(0)} \rangle 
+
\sum_{k=0,2,4} |k^{(0)} \rangle  
\frac{V_{k1}}{E^{(0)}_1-E^{(0)}_k}
+
\sum_{k,l \in S_1} 
|k^{(0)} \rangle  
\frac{V_{kl}V_{l1}}
{(E^{(0)}_1-E^{(0)}_k)(E^{(0)}_1-E^{(0)}_l)}
\end{eqnarray}
where 
$S_0$ represents $(l=1, k=2)$ or $(l=1, k=4)$ or 
$(l=3, k=2)$ or $(l=3, k=4)$ or $(l=3, k=6)$, and 
$S_1$ does $(l=0, k=3)$ or $(l=2, k=3)$ or 
$(l=2, k=5)$ or $(l=4, k=3)$ or $(l=4, k=5)$, or $(l=4, k=7)$.
Note that these eigenvectors are not normalized, so 
we need to renormalize them before or after calculations.

After some lengthy but straighforward calculations, 
we have 
\begin{eqnarray}
(q_S)_{10}
&=&
\langle 1| \bar{q}_S+b |0 \rangle
=(q_S)_{01}
=a(1+22 \varepsilon^2),
\label{eq:qs1}
\\
(q_S)_{00}
&=&
\langle 0| \bar{q}_S+b |0 \rangle
=b-6 a \varepsilon,
\\
(q_S)_{11}
&=&
\langle 1| \bar{q}_S+b |1 \rangle
=b- 18 a \varepsilon,
\\
(q^2_S)_{10}
&=&
\langle 1| (\bar{q}_S+b)^2 |0 \rangle
=(q^2_S)_{01}
=2 ab-20 a^2 \varepsilon +44 a b \varepsilon^2,
\\
(q^2_S)_{00}
&=&
\langle 0| (\bar{q}_S+b)^2 |0 \rangle
=a^2+b^2 -12 a b \varepsilon
 + 88 a^2 \varepsilon^2,
\\
(q^2_S)_{11}
&=&
\langle 1| (\bar{q}_S+b)^2 |1 \rangle
=3 a^2+b^2 
-36 ab \varepsilon
+568 a^2 \varepsilon^2
\label{eq:qs2}
\end{eqnarray}
where 
\begin{eqnarray}
a
=
\sqrt{\frac{\hbar}{2 \bar{\omega}_S}}
\end{eqnarray}
is the fundamental length charactering the system oscillator.

\section{The coefficients used in the formulas}
\label{app:coef}

Using the expression derived previously for the 
force-force correlation function \cite{FBS05a}, 
the coefficients in our VER and dephasing formulas 
are expressed as 
\begin{eqnarray}
\mathbf{C}^{\alpha \beta}  
&=&
\left(
\begin{array}{cc}
C^{\alpha \beta}_{--} & C^{\alpha \beta}_{+-} \\
C^{\alpha \beta}_{+-} & C^{\alpha \beta}_{++} \\
\end{array}
\right)
=
\left \{ 
(q_S)_{10} C_{S \alpha \beta} -(q^2_S)_{10} C_{SS \alpha \beta}
\right \}^2 
\mathbf{S}^{\alpha \beta},
\\
\mathbf{D}^{R \alpha \beta}  
&=&
\left(
\begin{array}{cc}
D^{R \alpha \beta}_{--} & D^{R\alpha \beta}_{+-} \\
D^{R \alpha \beta}_{+-} & D^{R \alpha \beta}_{++} \\
\end{array}
\right)
\nonumber
\\
&=&
\left \{
[(q_S)_{11}-(q_S)_{00}] C_{S \alpha \beta} -[(q^2_S)_{11}-(q^2_S)_{00}] C_{SS \alpha \beta}
\right \}^2 
\mathbf{S}^{\alpha \beta},
\\
\mathbf{D}^{I \alpha \beta}  
&=&
\left(
\begin{array}{cc}
D^{I \alpha \beta}_{--} & D^{I \alpha \beta}_{+-} \\
D^{I \alpha \beta}_{+-} & D^{I \alpha \beta}_{++} \\
\end{array}
\right)
\nonumber
\\
&=&
\left \{
[(q_S)_{11}-(q_S)_{00}] C_{S \alpha \beta} -[(q^2_S)_{11}-(q^2_S)_{00}] C_{SS \alpha \beta}
\right \} 
\nonumber
\\
&&
\times
\left \{
[(q_S)_{11}+(q_S)_{00}] C_{S \alpha \beta} -[(q^2_S)_{11}+(q^2_S)_{00}] C_{SS \alpha \beta}
\right \} 
\mathbf{S}^{\alpha \beta},
\\
\mathbf{E}^{\alpha \beta}  
&=&
\left(
\begin{array}{cc}
E^{\alpha \beta}_{--} & E^{\alpha \beta}_{+-} \\
E^{\alpha \beta}_{+-} & E^{\alpha \beta}_{++} \\
\end{array}
\right)
\nonumber
\\
&=&
\left \{
[(q_S)_{11}-(q_S)_{00}] C_{S \alpha \beta} -[(q^2_S)_{11}-(q^2_S)_{00}] C_{SS \alpha \beta}
\right \} 
\nonumber
\\
&&
\times
\left \{
(q_S)_{10} C_{S \alpha \beta} -(q^2_S)_{10} C_{SS \alpha \beta}
\right \} 
\mathbf{S}^{\alpha \beta},
\\
\mathbf{S}^{\alpha \beta}
&=&
\frac{\hbar^2}{2 \omega_{\alpha} \omega_{\beta}}
\left(
\begin{array}{cc}
(1+n_{\alpha})(1+n_{\beta}) & 2 (1+n_{\alpha})n_{\beta} \\ 
2 (1+n_{\alpha}) n_{\beta}  &  n_{\alpha} n_{\beta} \\
\end{array}
\right),
\\
\mathbf{C}^{\alpha}
&=&
\left(
\begin{array}{c}
C^{\alpha}_{-} \\
C^{\alpha}_{+} \\
\end{array}
\right)
=
(q_S^2)_{10}^2 C_{SS \alpha}^2 
\mathbf{R}^{\alpha},
\\
\mathbf{D}^{R\alpha}
&=&
\left(
\begin{array}{c}
D^{R\alpha}_{-} \\
D^{R\alpha}_{+} \\
\end{array}
\right)
=
[(q_S^2)_{11} - (q_S^2)_{00}]^2 
C_{SS \alpha}^2 
\mathbf{R}^{\alpha},
\\
\mathbf{D}^{I\alpha}
&=&
\left(
\begin{array}{c}
D^{I\alpha}_{-} \\
D^{I\alpha}_{+} \\
\end{array}
\right)
=
[(q_S^2)_{11} - (q_S^2)_{00}][(q_S^2)_{11} + (q_S^2)_{00}]
C_{SS \alpha}^2 
\mathbf{R}^{\alpha},
\\
\mathbf{E}^{\alpha}
&=&
\left(
\begin{array}{c}
E^{\alpha}_{-} \\
E^{\alpha}_{+} \\
\end{array}
\right)
=
[(q_S^2)_{11} - (q_S^2)_{00}](q_S^2)_{10}
C_{SS \alpha}^2 
\mathbf{R}^{\alpha},
\\
\mathbf{R}^{\alpha}
&=&
\frac{\hbar}{2 \omega_{\alpha}}
\left(
\begin{array}{c}
1 + n_{\alpha} \\
 n_{\alpha} \\
\end{array}
\right)
\end{eqnarray}
where $n_{\alpha}=1/(e^{\beta \hbar \omega_{\alpha}}-1)$ is 
the thermal phonon number.

To calculate $\bar{\omega}_S$ and $b$ in Eqs.~(\ref{eq:omegas}) and (\ref{eq:qs}),
we use the following
\begin{eqnarray}
\langle {\cal F}(t) \rangle
&=& \langle {\cal F}(0) \rangle
= \frac{\hbar}{2} \sum_{\alpha} 
\frac{C_{S \alpha \alpha}}{\omega_{\alpha}}
(1+ 2 n_{\alpha}), 
\\
\langle {\cal G}(t) \rangle
&=& \langle {\cal G}(0) \rangle
= \frac{\hbar}{2} \sum_{\alpha} 
\frac{C_{S S \alpha \alpha}}{\omega_{\alpha}}
(1+ 2 n_{\alpha}). 
\end{eqnarray}


\end{appendix}

\end{document}